\documentclass[%
 aip,
 amsmath,amssymb,
 reprint,%
]{revtex4-1}

\usepackage{graphicx}
\usepackage{dcolumn}
\usepackage{bm}

\usepackage[utf8]{inputenc}
\usepackage[T1]{fontenc}
\usepackage{mathptmx}

\usepackage{amsmath,amssymb,amsfonts}
\usepackage{algorithmic}
\usepackage{textcomp}
\usepackage{multirow}
\usepackage{soul}
\usepackage[caption=false,font=footnotesize,subrefformat=parens,labelformat=parens]{subfig}

\begin{document}

\preprint{AIP/123-QED}

\title{Radio frequency mixing modules for superconducting qubit room temperature control systems}

\author{Yilun Xu}
\author{Gang Huang}%
 \email{ghuang@lbl.gov}
\author{David I. Santiago}
\author{Irfan Siddiqi}
\affiliation{ 
Lawrence Berkeley National Laboratory, Berkeley, CA 94720, USA
}%


\begin{abstract}
As the number of qubits in nascent quantum processing units increases, the connectorized RF (radio frequency) analog circuits used in first generation experiments become exceedingly complex. 
The physical size, cost and electrical failure rate all become limiting factors in the extensibility of control systems. 
We have developed a series of compact RF mixing boards to address this challenge by integrating I/Q quadrature mixing, IF(intermediate frequency)/LO(local oscillator)/RF power level adjustments, and DC (direct current) bias fine tuning on a 40~mm $\times $ 80~mm 4-layer PCB (printed circuit board) board with EMI (electromagnetic interference) shielding.
The RF mixing module is designed to work with RF and LO frequencies between 2.5 and 8.5~GHz. 
The typical image rejection and adjacent channel isolation are measured to be $\sim$27~dBc and $\sim$50~dB. 
By scanning the drive phase in a loopback test, the module short-term amplitude and phase linearity are typically measured to be 5$\times$10$^{-4}$ (V$_{\mathrm{pp}}$/V$_{\mathrm{mean}}$) and 1$\times$10$^{-3}$~radian (pk-pk). 
The operation of RF mixing board was validated by integrating it into the room temperature control system of a superconducting quantum processor and executing randomized benchmarking characterization of single and two qubit gates. 
We measured a single-qubit process infidelity of $9.3(3) \times 10^{-4}$ and a two-qubit process infidelity of $2.7(1) \times 10^{-2}$.
\end{abstract}

\maketitle

\section{Introduction}
\label{sec:introduction}
Quantum computing has the potential to solve computational problems that are classically intractable in the post-Moore era \cite{preskill2018quantum,arute2019quantum}. 
The superconducting quantum bit (qubit) is one of the most promising candidates that have demonstrated basic quantum computing functionality \cite{walter2017rapid,heinsoo2018rapid}.

\begin{figure}[t!]
\centering
\includegraphics[width=0.8\linewidth]{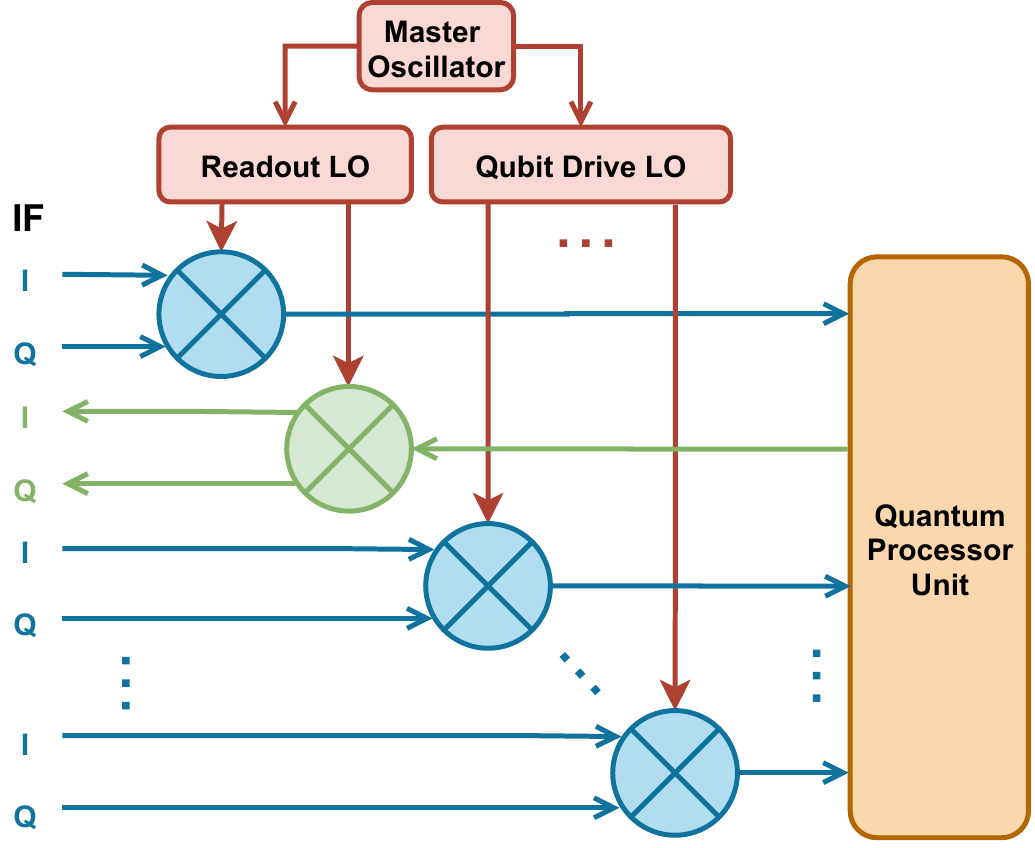}
\caption{The up converter mixes IF signals with an LO to generate the RF signals needed to control logical operations in the quantum processor unit, while the down converter mixes down readout signals with the same LO, generating the quantum-state-dependent IF signals. The up converter (blue) and the down converter (green) act as the transmitter and the receiver, respectively.}
\label{fig1}
\end{figure}

High resolution, low noise RF (radio frequency) signals, which are typically in the 4--8 GHz frequency range with several hundreds of MHz bandwidth, are used to control and measure superconducting qubits \cite{zurich2020frequency,ryan2017hardware,perego2018scalable}. 
These signals can be generated and detected using the standard heterodyne technique as shown in Fig.~\ref{fig1}. 
The intermediate frequency (IF) signals are generated or digitized by way of a digital-to-analog converter (DAC) or analog-to-digital converter (ADC), respectively. 
Employing a limited IF bandwidth allows the system to use lower frequency ADCs and DACs for better noise performance and lower cost. 
The IF signals are shifted up or down in frequency by the RF mixing module and the low noise local oscillator (LO) \cite{vainsencher2019superconducting}. 

Alternatively, a direct digital synthesis module can synthesize microwave waveforms in the GHz regime which can potentially simplify the system. 
However, current direct RF synthesis devices are still limited by their signal resolution, frequency range, cost, and usability \cite{raftery2017direct}.

In traditional experimental systems, connectorized RF components are widely used because they are easy to assemble and interchange \cite{salathe2018low}. 
However, each connector is bulky and can be a potential failure point in the long run, especially in extant systems.
Based on the topology of a connectorized component configuration, we developed a series of compact RF mixing modules to reduce both cost and size, as well as improve system reliability. 
Such integration and optimization will help room-temperature-based control system keep pace with advancements in quantum information processor complexity and scale. 

In this work, we developed RF mixing modules \cite{gitrepo2021} and tested them with the superconducting quantum processors operating in the Quantum Nanoelectronics Laboratory (QNL) at the University of California, Berkeley. 
We compared the output of standard qubit benchmarks executed using our next generation electronics with standard connectorized modules. 
The solution proposed here exhibits low-noise, high-reliability operation and is now becoming the laboratory standard for microwave frequency modulation/demodulation across many different experimental configurations. 

\section{RF mixing module development}
Specifically, we developed three RF mixing modules: one down converter module (DOWN: DN) and two up converter modules (UP: UPH, UPL) with different output power levels. The UPH and UPL labels correspondingly indicate the up converter module with higher and lower output power.
The UP variants differ by an additional amplifier in the RF channel to accommodate varying degrees of cryogenic attenuation associated with the different RF cable lengths and materials.
For design simplicity, we start with one board per qubit and then study the crosstalk between boards.

\begin{figure}[t!]
\centering
\subfloat[]{\includegraphics[width=0.8\linewidth]{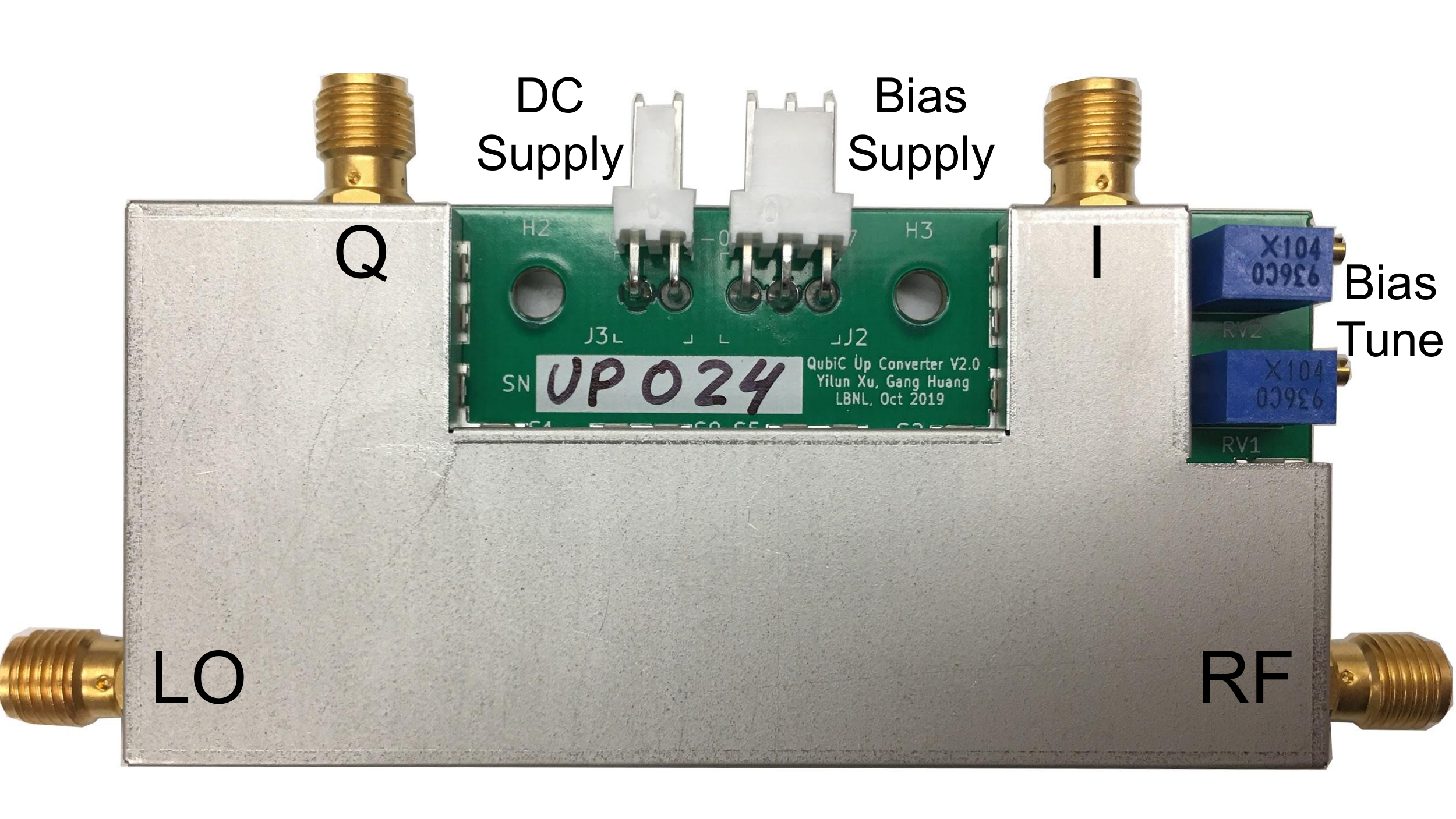}
\label{fig2a}}
\vfil
\subfloat[]{\includegraphics[width=0.8\linewidth]{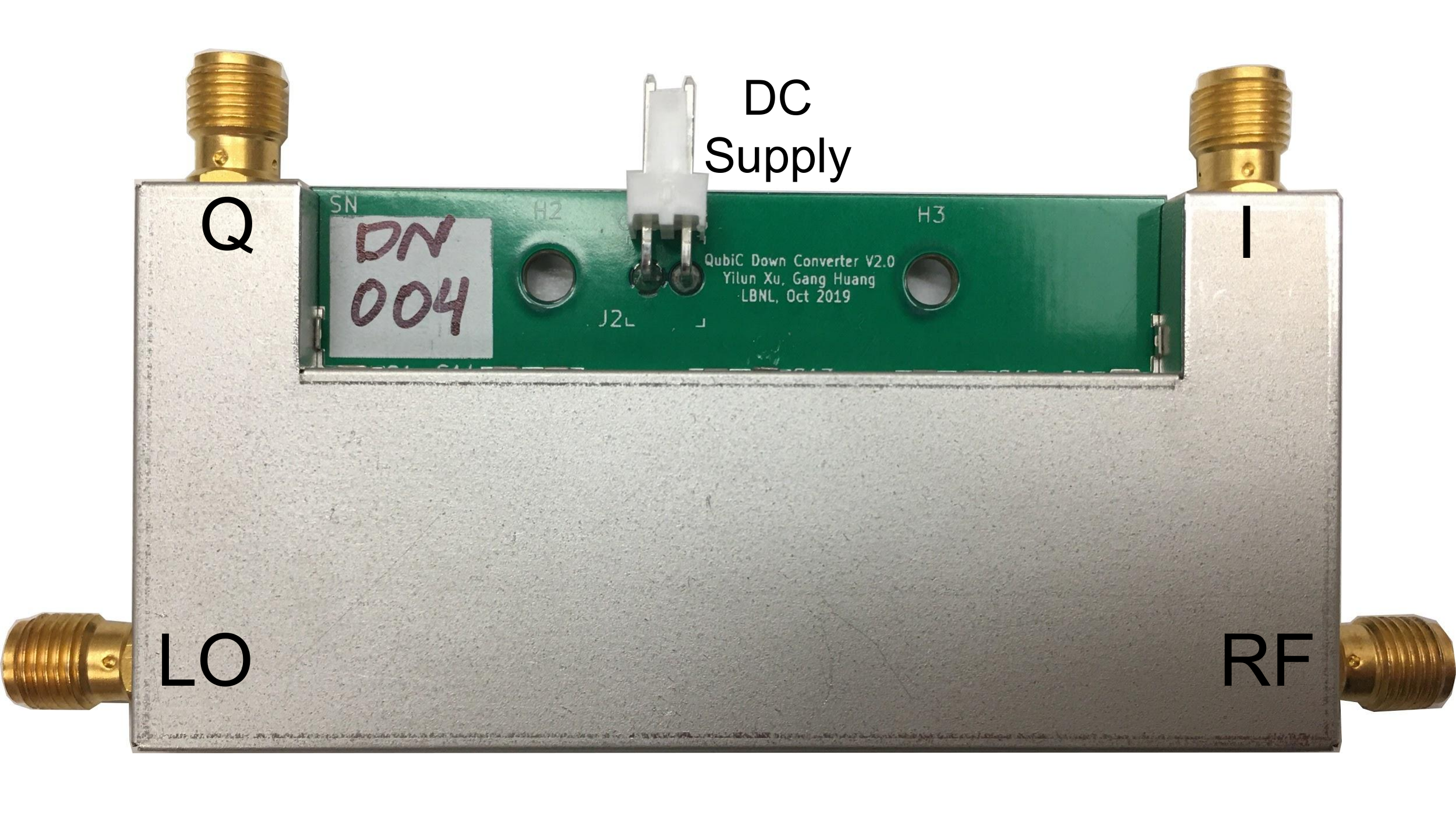}
\label{fig2b}}
\caption{RF mixing boards. (a) Up converter with EMI shielding. The input SMA (SubMiniature version A) ports are ``LO'', ``I'' and ``Q'', while the output SMA port is ``RF''. The 2-pin DC (direct current) connector is for the power supply, and the 3-pin DC connector is for the bias supply. Two potentiometers (for I and Q channels) can be used to tune the bias and null the LO signal. (b) Down converter with EMI shielding. The input SMA ports are ``LO'' and ``RF'', while the output SMA ports are ``I'' and ``Q''. The 2-pin DC connector is for the power supply.}
\label{fig2}
\end{figure}

The RF mixing module size should be as small as possible to support an increasing number of qubits. 
As shown in Fig.~\ref{fig2}, the actual RF mixing bare board (up converter or down converter) size is 40~mm $\times $ 80~mm. 
The total number of RF mixing module channels scales linearly with the number of qubits, as does the cost of boards. 
The RF mixing module should also be designed with value engineering.

Every room temperature component has thermal noise, and active components contribute additional flicker and other noise to the system. 
Undesired signals can also come from adjacent channel crosstalk or environmental electromagnetic interference (EMI) along the whole signal transmission line. 
The noise performance of the RF mixing module directly affects qubit performance. 
Therefore, the module needs to be implemented with low noise components. 
Typically higher signal power level can improve signal-to-noise ratio (SNR), albeit at the expense of nonlinear effects in amplifiers or mixers. 
Thus, the signal power level needs to be designed to reach an optimal compromise between signal-to-noise and linearity.

\subsection{Modules design}
\begin{figure}[t!]
\centering
\subfloat[]{\includegraphics[width=1.0\linewidth]{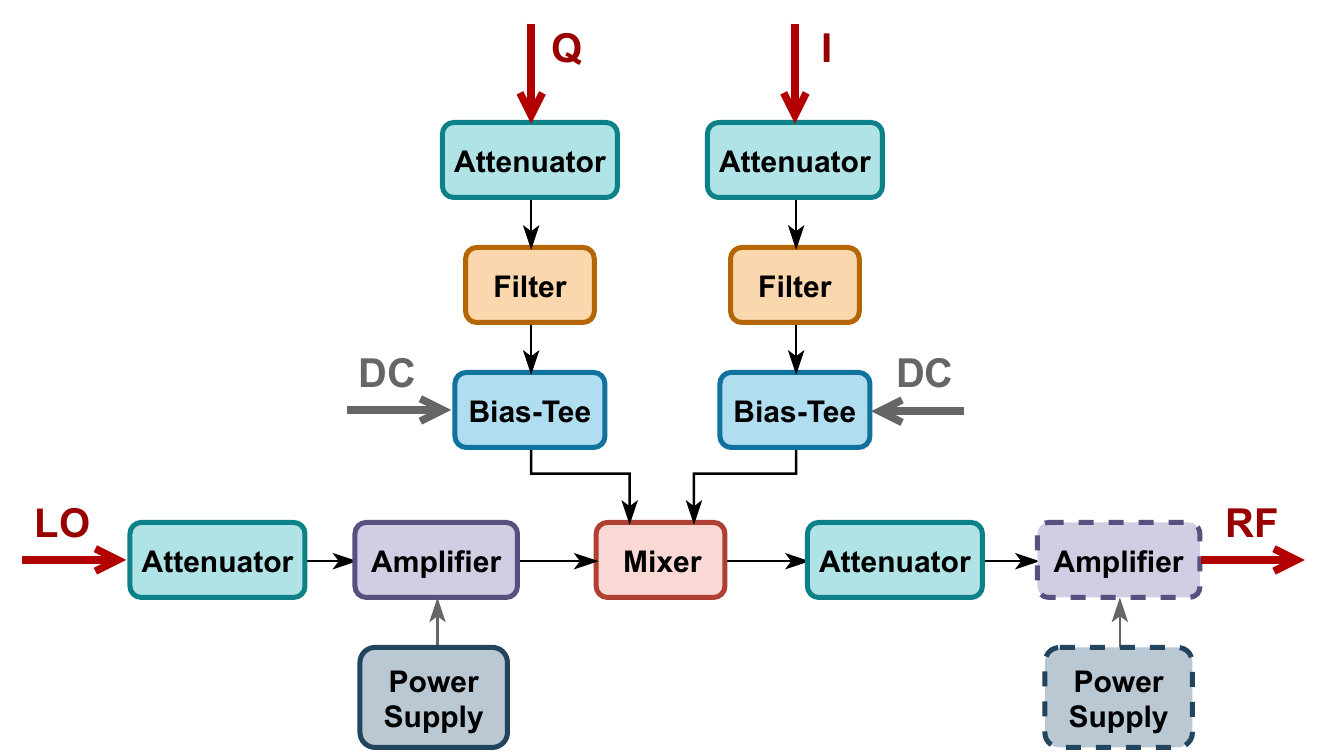}
\label{fig3a}}
\vfil
\subfloat[]{\includegraphics[width=1.0\linewidth]{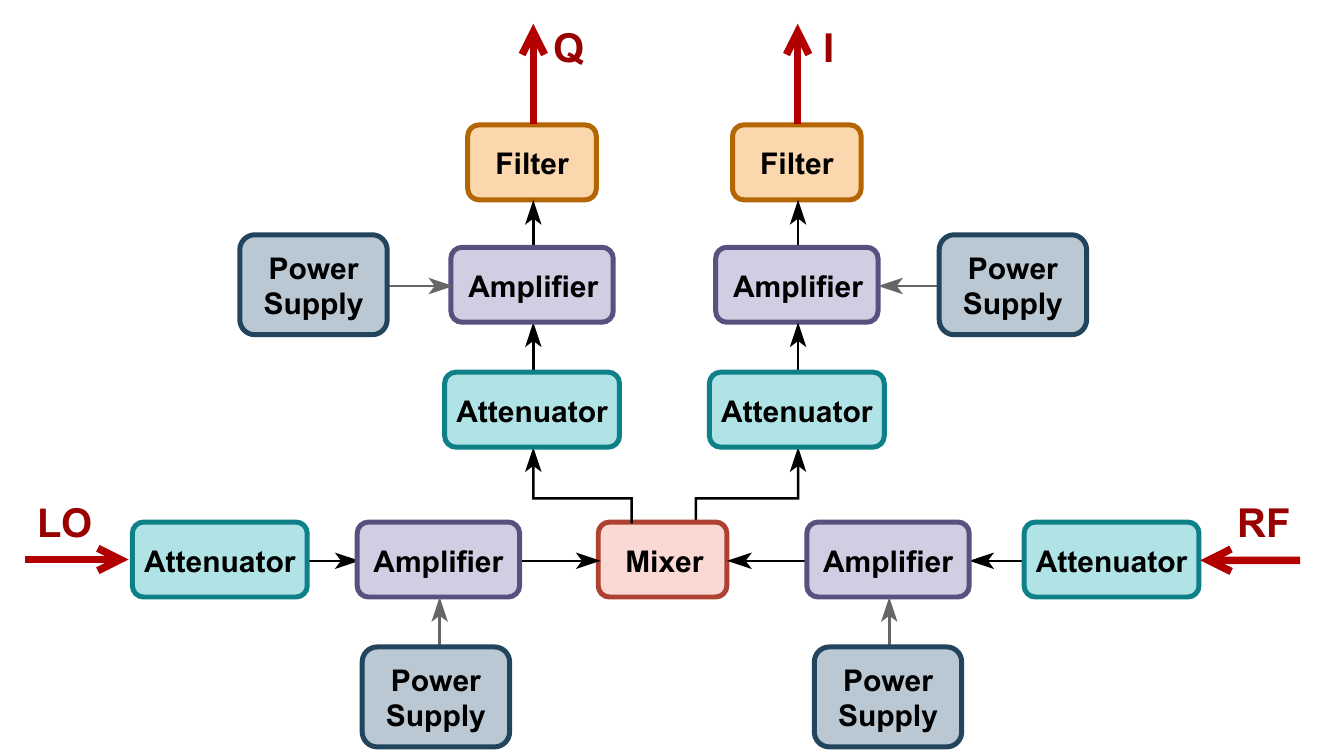}
\label{fig3b}}
\caption{Block diagram of RF mixing modules. (a) Up converter. We have variations with/without an amplifier in the RF channel for the UPH/UPL modules. (b) Down converter.}
\label{fig3}
\end{figure}

All three RF mixing modules are comprised of several interconnected submodules as shown in Fig.~\ref{fig3}. 
Each submodule was designed independently and can be repurposed as a modular component to implement the overall module functionalities as needed. 

\subsubsection{Mixer}
The IQ mixer is used to shift the qubit control/readout frequency from/to the IF which can be easily handled by the ADC/DACs.
The mixer selection needs to cover the required frequency range in the superconducting qubit setup. 
Most 5G communication transceivers available on the market cut off at $\sim$6~GHz and are not applicable for qubit control. 
The passive IQ mixer HMC8193 from Analog Devices covers an RF and LO range from 2.5~GHz to 8.5~GHz, and an IF range from DC to 4~GHz \cite{hmc8193}, thus meeting our requirement for UP and DN modules. 

\subsubsection{Power level}
The amplifiers and attenuators are designed to adjust the power level to meet the gate pulse requirements, adjust the signal-to-noise ratio at different locations, and also to provide attenuation to suppress mismatch signals bouncing between discontinuities. 

The LO signal is distributed from the LO generation chassis and is fanned out to multiple UP/DN modules. 
It is convenient to have an independent LO amplifier together with an attenuator for each channel to provide the required power for the mixer, while also increasing the isolation between channels. 
At 8~GHz, the mini-circuits CMA-83LN+ provides $\sim$20~dB gain in the band, while its noise figure is 1.8~dB and the P1dB is 16~dBm \cite{cma83ln}. 
The same amplifier is also used in the RF channel. 

Qubit readout resonator signals are typically amplified in the dilution refrigerator by a cryogenic, superconducting traveling wave parametric amplifier (TWPA) \cite{macklin2015quantum} and/or high-electron-mobility transistor (HEMT) to increase SNR. 
After down conversion at room temperature, the IF signal level is too low for ADC to directly digitize with enough resolution. 
A low noise IF amplifier after the mixer is needed to boost the voltage while minimally degrading SNR. 
The mini-circuits LHA-13LN+ amplifier is employed as it can provide $\sim$22~dB gain in band, with a low noise figure of 1.1~dB across the band \cite{lha13ln}.

Susumu PAT0510S series attenuators are used in the RF mixing modules, as they have $<$1.3~VSWR (voltage standing wave ratio), good attenuation flatness across the band, the compact footprint, and are low cost \cite{pat0510s}.

\subsubsection{Bias-Tee}
A undesirable signal in the up conversion is LO leakage through the mixer which can degrade qubit performace.
This deleterious signal can be suppressed by precisely adjusting the DC bias level on the mixer I (in-phase) and Q (quadrature) ports. 
To combine the DC bias and RF signal, we choose the Mini-Circuits TCBT-14R+ as a bias-tee which provides an ultra-wide frequency range with excellent isolation performance over the band.
We also added two potentiometers to the up converter to tune the bias and null the LO. 
Furthermore, each potentiometer can be replaced by a jumper to allow for the automatic control of DC biases to compensate for the drift in the nulling signal.

\subsubsection{Low noise power supply}
Each amplifier is powered by a separate low dropout regulator (LDO) to minimize the potential crosstalk through the power line.
All the LDOs share the same off board switcher regulator. 
Switcher supplies provides high efficiency in converting voltages close to the required amplifier voltage. 
A LDO further regulates the bias to the final voltage and rejects any switching noise from the switcher. 
Texas Instruments LP38798 is used in the RF mixing module, providing ultra-low output noise at 5$\mu$V (rms) from 10~Hz to 100~kHz and high power supply rejection ratio \cite{lp38798}.
Both the up and the down converter operates on a 6.3~V supply, while the up converter also uses an additional $\pm$1.8~V supply for LO nulling. 

\subsubsection{EMI shielding and thermal design}
Electromagnetic interference reduces electronic signal integrity and can limit the overall system performance.
EMI shielding is implemented on the RF mixing boards to protect the sensitive signal from disturbances in the external electromagnetic environment, and also to prevent the signal from leaking out and interfering with surrounding electronics. 
On the RF mixing boards, EMI shielding clips and custom-shaped covers are located around the RF sections of the boards as well as the SMA connector.

The back side of the board is designed to be a flat ground plane with no components to help with EMI. 
This flat back design also provides a simple way to attach an external heat sink or heat mess to cool or stabilize the board temperature.

\subsubsection{Impedance control and board stack-up}
The impedance is nominally designed to be 50~$\Omega$ across the board. 
In practice,  however, values are impacted by variations in the the trace impedance, footprints of components, and the SMA connectors themselves. 
The trace impedance is determined by the PCB (printed circuit board) process and the materials from the PCB vendor. 
We evaluated several SMA connectors and their footprints on different materials with the impedance control service from the PCB vendor. 
The impedance of the RF mixing board was measured to be 47~$\Omega$ by time-domain reflectometry (TDR) after assembly with SMA connectors (Cinch Connectivity Solutions Johnson 142-0771-831). 
The insertion loss was measured by the determining the S21 of a coupon board of similar length in the $<$10~GHz frequency band. 
We also compared the performance of Rogers (4003C) and FR4 substrates with respect to insertion loss in the band, and measured $\sim$2~dB power loss across the board at 10~GHz. 
Since we have enough output power, we decided to use FR4 to reduce cost. 

Both the UP and DN modules are designed to be 4-layer PCB boards: signal layer, ground layer, power layer and ground layer from top to bottom. 
This stack-up design can effectively reduce electromagnetic emissions and crosstalk, and improve the signal integrity to provide a low inductance power distribution network. 
Besides the 2 ground layers, the large pieces of copper on the top signal layer cover the area without components and traces, thereby also connecting to the ground plane. 
Ground planes on different layers are connected through vias, which help reduce ground loops and keep the length of return loops short in the system.

\section{RF mixing module test}
\subsection{Bench test -- RF specification}
The RF mixing board specifications are summarized in Table.\ref{tab1}. The board faithfully maintains the excellent performance from the monolithic microwave integrated circuit (MMIC) mixer HMC8193. 


\begin{table}
\centering
\subfloat[Common specifications.\label{tab1a}]{
\centering
\setlength{\tabcolsep}{3pt}
\begin{tabular}{l|ccc|l}
\hline
Parameter & Min & Typ & Max & Unit \\
\hline
Radio Frequency & 2.5 &  & 8.5 & GHz \\
\hline
Local Oscillator & & & & \\ 
\quad Frequency & 2.5 &  & 8.5 & GHz \\
\quad Drive Level &  & 0 &  & dBm \\
\hline
Intermediate Frequency & DC &  & 0.5 & GHz \\
\hline
Return Loss & & & & \\
\quad RF & 3.5 & 8.5 & 24 & dB \\
\quad LO & 4 & 8.5 & 21 & dB \\
\quad IF & 9.5 & 13.5 & 20 & dB \\
\hline
Isolation $^{\mathrm{a}}$ & & & & \\
\quad RF to IF & 21.5 & 49.5 &  & dB \\
\quad LO to RF & 40.5 & 51.5 &  & dB \\
\quad LO to IF & 25 & 50.5 &  & dB \\
\hline
\multicolumn{3}{l}{$^{\mathrm{a}}$Down converter performance.}
\end{tabular}}
\vfil
\subfloat[Typical specifications for DN (down converter), UPH (upconverter with additional amplifier), and UPL (up converter without additional amplifier).\label{tab1b}]{
\centering
\setlength{\tabcolsep}{3pt}
\begin{tabular}{l|c|c|c|l}
\hline
Parameter & DN & UPH & UPL & Unit \\
\hline
Drive Level & $<$-27 & $<$4 & $<$4 & dBm \\
Conversion Gain & 31 & 7 & -13 & dB \\
Image/Sideband Rejection $^{\mathrm{b}}$ & 27 & 27.5 & 26.5 & dBc \\
Input Third-Order Intercept & -3.5 & 21 &  & dBm \\
Input Second-Order Intercept & 8 &  &  & dBm \\
Input 1 dB Compression Point & -13.5 &  &  & dBm \\
Operating Voltage & 6.3 & 6.3 & 6.3 & V \\
Operating Current & 410 & 140 & 70 & mA \\
\hline
\multicolumn{5}{l}{$^{\mathrm{b}}$Image rejection for DN, sideband rejection for UPH, UPL.}
\end{tabular}}
\caption{Specifications of RF mixing modules.}
\label{tab1}
\end{table}

The RF, LO and IF pins are AC-coupled and matched to be 50~$\Omega$. 
For the current application, the low pass filter in the IF channel cuts off the IF range to 500~MHz.
However, the mixer offers a wide IF range of DC to 4~GHz, so it is easy to expand the IF range by replacing the low pass filter. 
Since an internal amplifier is included in the LO path, only 0~dBm power is needed for the LO drive level of the UP/DN module. 
The RF mixing board offers 27~dBc image rejection and thereby eliminates the need for filtering of unwanted sidebands. 
Amplifiers and attenuators are cascaded in the RF/LO/IF path to enhance the isolation ($\sim$50~dB) and suppress crosstalk. 

\subsection{Bench test -- Loopback noise}
\begin{figure}[t!]
\centering
\includegraphics[width=1.0\linewidth]{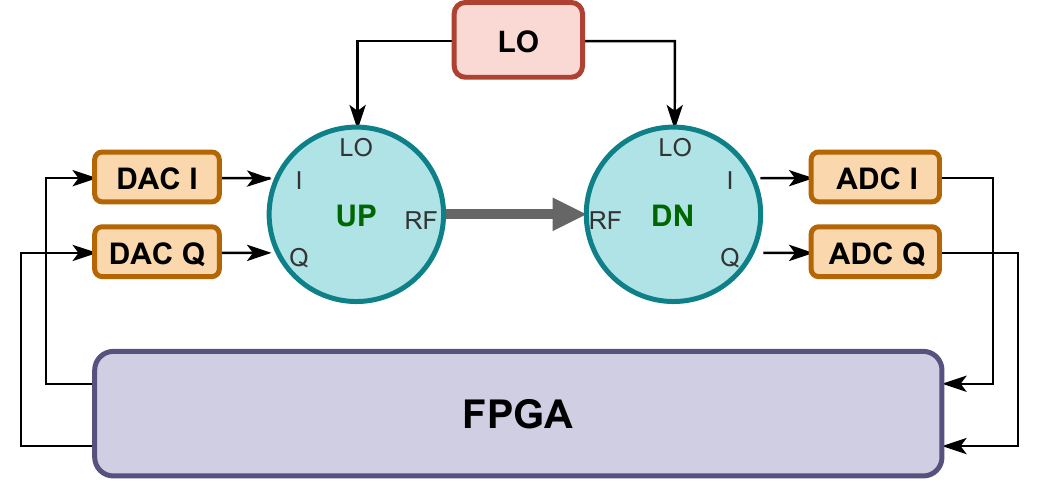}
\caption{Loopback test setup. The I/Q signals were generated by the FPGA, then fed into the DAC pairs. The UP mixed up the I/Q signals to an RF signal. The RF signal was directly connected back to the DN for down mixing, followed by the ADC digitization and the FPGA process. The UP and DN shared the same LO. We scanned the phase of the transmitted I/Q signal in some selected frequencies, and observed the amplitude and phase responses of the receiver integrated complex signal.}
\label{fig4}
\end{figure}

To measure the noise performance of the RF mixing UP/DN modules, we used a QubiC (Qubit Control) prototype chassis \cite{huang2020qubic} with a set of UP/DN converters to implement a loopback test as shown in Fig.~\ref{fig4}. 
The QubiC system is an FPGA (field-programmable gate array) based Qubit Control system, developed at Lawrence Berkeley National Laboratory \cite{xu2020qubic}. 
A QubiC prototype chassis provides 8 channels 16 bit DACs and 8 channels 16 bit ADCs at 1~GSPS (giga samples per second).

\begin{figure}[t!]
\centering
\subfloat[]{\includegraphics[width=0.8\linewidth]{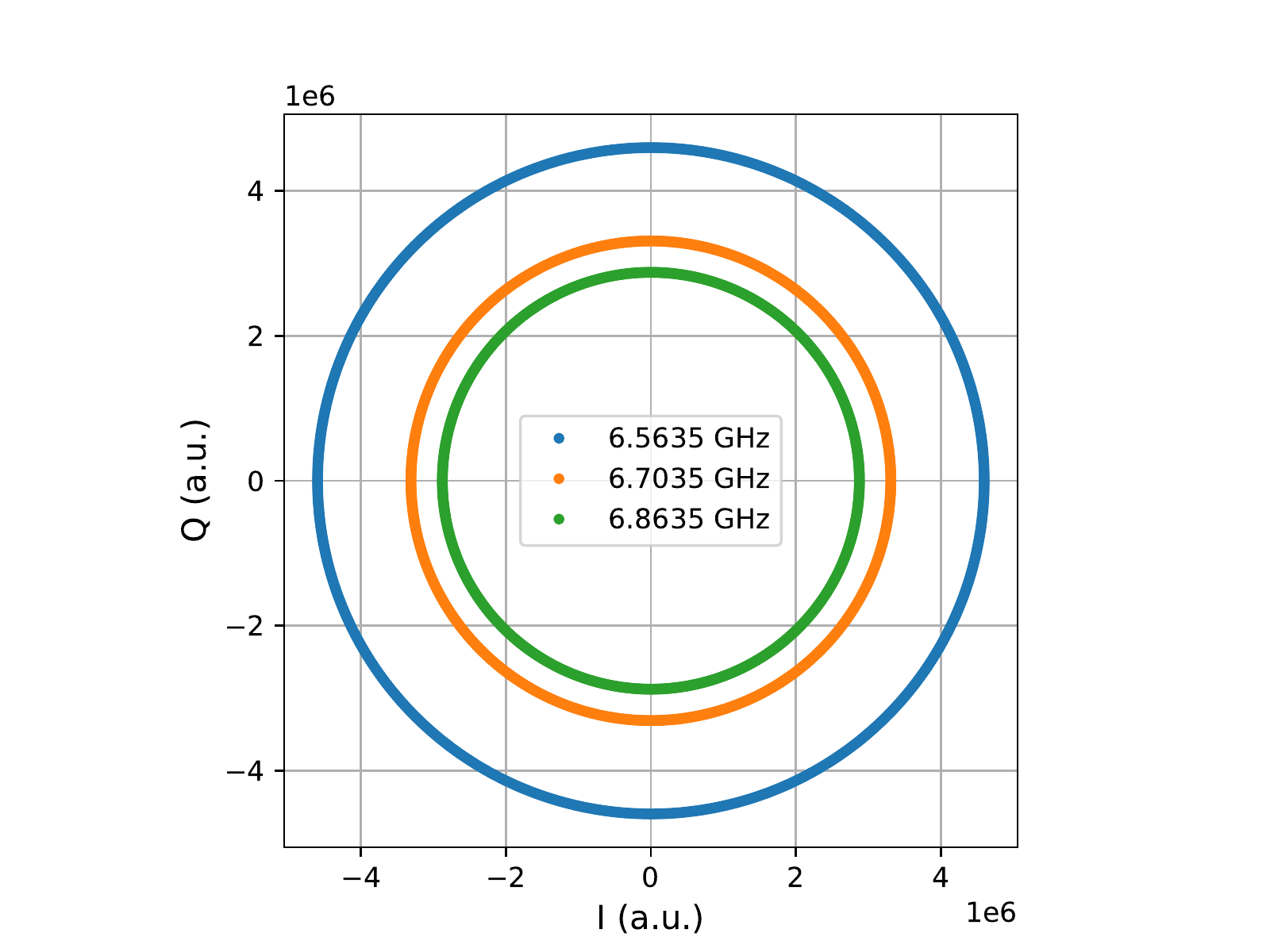}
\label{fig5a}}
\vfil
\subfloat[]{\includegraphics[width=0.8\linewidth]{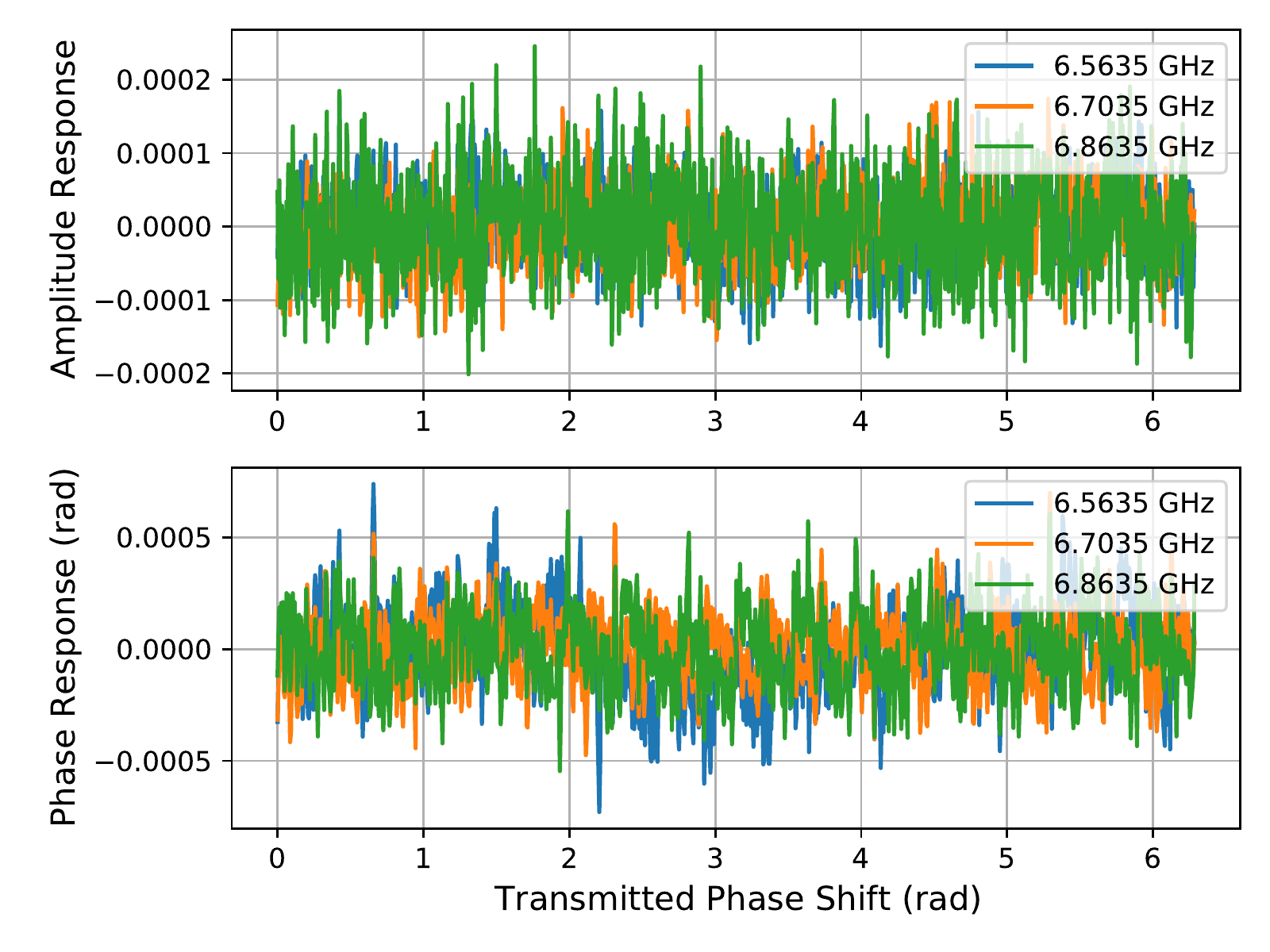}
\label{fig5b}}
\caption{Loopback test results. (a) IQ plot of the receiver integrated complex signal, obtained while scanning the phase of the transmitted I/Q signal over a 2$\pi$ cycle. (b) The amplitude response (V$_{\mathrm{pp}}$/V$_{\mathrm{mean}}$) and the phase response (detrended) of the receiver integrated complex signal, obtained while scanning the phase of the transmitted I/Q signal over a 2$\pi$ cycle.}
\label{fig5}
\end{figure}

The QubiC FPGA and DACs generates I and Q components at IF. 
The transmitted I/Q signals form a complex signal which contains the amplitude and phase information.
The UP module converts the I/Q signals from IF to RF, which is directly transmitted to the RF port of the DN module. 
On the receiver side, it is a two-stage mixing process, which is an analog mixing followed by a digital mixing.
The first mixing is that the DN module receives the RF signal and converts it back to I/Q signals at IF.
The second mixing is that the ADC digitizes the IF signal and the FPGA further down converts it to the baseband by digital mixing with the digital local oscillator (DLO). 
The baseband I and Q values are integrated by the accumulator and stored in the accumulator buffer, which is a complex signal and can be readout later.
We can easily extract the amplitude and phase from this receiver integrated complex signal. 
To evaluate the noise performance of the UP and DN modules, we can measure the amplitude and phase responses of the integrated complex signal, while scanning the phase of the transmitted I/Q signal over the 2$\pi$ period. 
We chose three different drive frequencies (6.5635~GHz, 6.7035~GHz, 6.8635~GHz) as the example for this test. 
As shown in Fig.~\ref{fig5a}, the IQ curve is close to a circle, which indicates good amplitude and phase balance. 

To quantify the noise performance, we defined the peakpeak-to-mean ratio (V$_{\mathrm{pp}}$/V$_{\mathrm{mean}}$) for the ``amplitude'' and the detrended phase for the ``phase''. 
The receiver integrated complex signal has homogeneous amplitude and phase noise performance, while scanning the whole phase cycle of the transmitted I/Q signal, as shown in Fig.~\ref{fig5b}. 
The linearity of the amplitude response and phase response is summarized in Table.\ref{tab2}. 
Compared to the widely-used commercial Marki mixers \cite{mliq0416}, the RF mixing module has better amplitude performance and comparable phase performance.

\begin{table}
\centering
\caption{Amplitude linearity and phase linearity of different mixing schemes.}
\label{table}
\setlength{\tabcolsep}{3pt}
\begin{tabular}{c|c|c|c|c}
\hline
\multirow{3}{*}{Freq (GHz)}&\multicolumn{2}{c|}{RF Mixing Module}&\multicolumn{2}{c}{Marki MLIQ0416L}\\
\cline{2-5}
& Amp & Phase & Amp & Phase \\
& (V$_{\mathrm{pp}}$/V$_{\mathrm{mean}}$) & (rad) & (V$_{\mathrm{pp}}$/V$_{\mathrm{mean}}$) & (rad) \\
\hline
6.5635 & 3$\times$10$^{-4}$ & 1$\times$10$^{-3}$ & 9$\times$10$^{-4}$ & 1$\times$10$^{-3}$ \\ 
6.7035 & 3$\times$10$^{-4}$ & 1$\times$10$^{-3}$ & 5$\times$10$^{-4}$ & 1$\times$10$^{-3}$ \\
6.8635 & 4$\times$10$^{-4}$ & 1$\times$10$^{-3}$ & 4$\times$10$^{-4}$ & 1$\times$10$^{-3}$ \\
\hline
\end{tabular}
\label{tab2}
\end{table}

\begin{figure}[t!]
\centering
\includegraphics[width=1.0\linewidth]{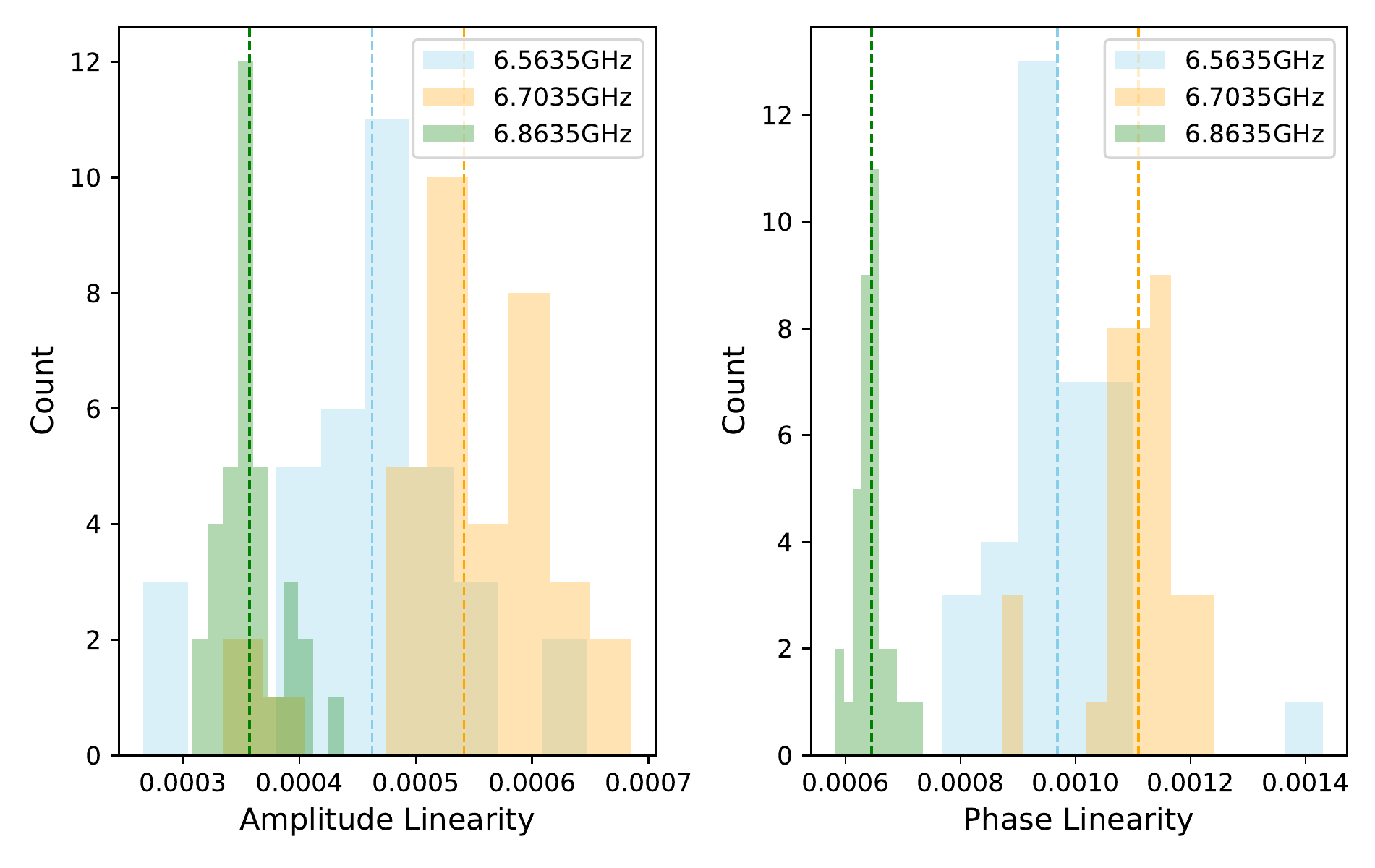}
\caption{Linearity statistics of RF mixing modules. 35 RF mixing modules were tested in three different drive frequencies.}
\label{fig6}
\end{figure}

We tested a series of RF mixing modules for statistics, yielding a high pass rate on amplitude and phase linearity. 
As shown in Fig.~\ref{fig6}, the short-term amplitude and phase linearity of the RF mixing modules are typically measured to be 5$\times$10$^{-4}$ (V$_{\mathrm{pp}}$/V$_{\mathrm{mean}}$) and 1$\times$10$^{-3}$~radian (pk-pk).

\subsection{Test on quantum hardware}
\begin{figure}[t!]
\centering
\subfloat[]{\includegraphics[width=1.0\linewidth]{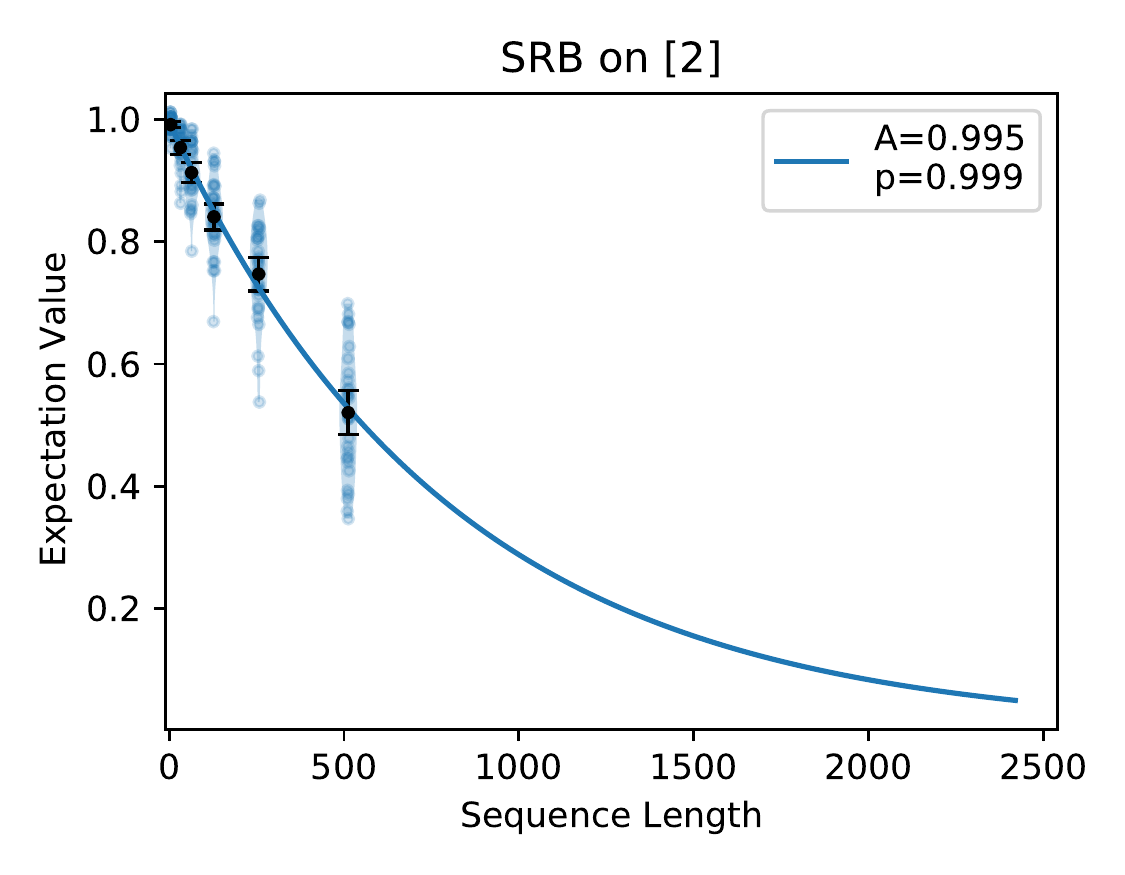}
\label{fig7a}}
\vfil
\subfloat[]{\includegraphics[width=1.0\linewidth]{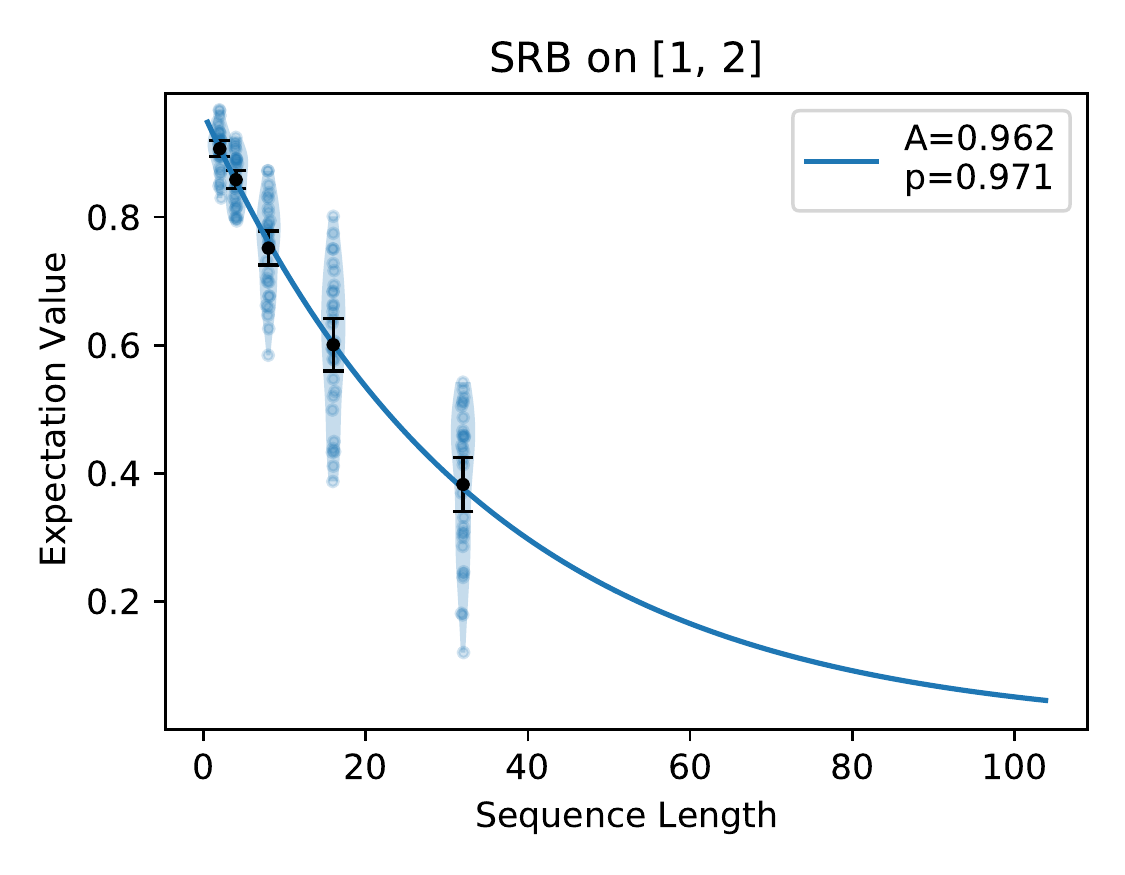}
\label{fig7b}}
\caption{Streamlined randomized benchmarking results. The RB data acquired by QubiC are fed to the True-Q software for fitting. The curves are fitted with the function of $\mathrm{A}\mathrm{p}^\mathrm{m}$. The sequence length $\mathrm{m}$ is expressed in terms of Clifford gates. (a) Single-qubit RB. (b) Two-qubit RB.}
\label{fig7}
\end{figure}

Randomized benchmarking (RB) is a widely-used randomization method for benchmarking quantum gates, which is not sensitive to the state preparation and measurement errors \cite{knill2008randomized}. 
In order to validate the functionality of RF mixing modules on quantum hardware, we integrated them with the QubiC system to perform the RB measurement on a 8-qubit quantum processor \cite{kreikebaum2020improving} residing in the dilution refrigerator at the Quantum Nanoelectronics Laboratory (QNL) in University of California, Berkeley. 
Using our automatic single qubit characterization \cite{xu2020automatic} and two-qubit gate optimization \cite{xu2021automatic} protocols, we obtained single-qubit and two-qubit RB results.
The process infidelity, which is the total probability that an error acts on the targeted systems during a random gate, can be extracted from the RB data \cite{magesan2011scalable}.
The streamlined RB (SRB) \cite{trueq2021srb} is one of the RB protocols widely adopted by the community to characterize the process infidelity. 
We implemented the SRB protocol by using the True-Q software \cite{beale_stefanie_j_2020_3945250} to generate the randomized circuits and fit the data. 
The qubit control and data acquisition were realized by the QubiC system together with the RF mixing modules under test. 
As shown in Fig.~\ref{fig7}, the single-qubit process infidelity was measured to be $9.3(3) \times 10^{-4}$, while the two-qubit process infidelity was measured to be $2.7(1) \times 10^{-2}$. 

The RB results demonstrated that the RF mixing modules are capable of delivering high fidelity gates on state of the art device.
We believe that the RF mixing module is not the current limiting factor for higher fidelity.
The RF mixing modules have been routinely employed in QNL and Advanced Quantum Testbed (AQT) experimental activities.

\section{Conclusion}
We developed a series of compact RF mixing modules for superconducting qubit room temperature control systems. 
This is a start point to overcome challenges in extensibility arising from connectorized analog front-end components for superconducting quantum processor control. 
The RF mixing module is designed to operate in the 2.5 to 8.5~GHz RF and LO range, and offers 27~dBc (typical) image rejection and $\sim$50~dB (typical) isolation between channels. 
The loopback test showed the RF mixing module reached 5$\times$10$^{-4}$ (V$_{\mathrm{pp}}$/V$_{\mathrm{mean}}$) amplitude linearity and 1$\times$10$^{-3}$~radian (pk-pk) phase linearity over the whole phase cycle. 
Using the QubiC system assembled with this RF mixing module, we tested the operation of a superconducting quantum processor in a dilution refrigerator and measured $9.3(3) \times 10^{-4}$ (single-qubit) and $2.7(1) \times 10^{-2}$ (two-qubit) process infidelities from randomized benchmarking characterization.
This confirmed that the functionality and performance of the RF mixing modules met the desired qubit control requirements.

\begin{acknowledgments}
This work was supported by the High Energy Physics QUANTISED program, the Advanced Scientific Computing Research Testbeds for Science program, and the Quantum Systems Accelerator under the Office of Science of the U.S. Department of Energy under Contract No. DE-AC02-05CH11231. 
The authors would like to thank Ravi Naik, Jean-Loup Ville, Bradley Mitchell, Jie Luo, and Brian Marinelli from the University of California, Berkeley; Alexis Morvan, Jan Balewski, Wim Lavrijsen, and Kasra Nowrouzi from the Lawrence Berkeley National Laboratory; and Quantum Benchmark,Inc., for their support.
\end{acknowledgments}

\section*{Data Availability Statement}
The data that support the findings of this study are available from the corresponding author upon reasonable request.

\nocite{*}
\bibliography{aipsamp}

\end{document}